\documentclass[pdftex,11pt,a4paper,parskip=half,headsepline]{scrartcl}

\usepackage[english]{babel}
\usepackage[singlespacing]{setspace}
\usepackage[automark]{scrlayer-scrpage}
\usepackage[utf8]{inputenc}
\usepackage[T1]{fontenc}
\usepackage{amsmath}
\usepackage{amssymb}
\usepackage{amsthm}
\usepackage{graphicx}
\usepackage{xcolor}
\usepackage{longtable}
\usepackage[longtable]{multirow} 
\usepackage{booktabs} 
\usepackage{url}
\usepackage{hyperref}
\usepackage{todonotes}
\usepackage{titling}  

\usepackage[backend=biber, bibstyle=apa, maxcitenames=2, citestyle=apa, isbn=false, doi=true, url=true, eprint=false]{biblatex}
\AtEveryBibitem{\clearfield{month}}
\AtEveryCitekey{\clearfield{month}}
\AtEveryBibitem{\clearfield{day}}
\AtEveryCitekey{\clearfield{day}}
\AtEveryBibitem{\clearfield{note}}
\AtEveryCitekey{\clearfield{note}}
\NewBibliographyString{toappear}
\DefineBibliographyStrings{english}{%
  toappear = {to appear in:},
  version = {R package version},
} 
\usepackage{csquotes}

\graphicspath{ {./figures/} }

\title{Why is Regularization Underused? An Empirical Study on Trust and Adoption of Statistical Methods}

\author{Konstantin Emil Thiel$^{*,\dagger,1,2}$, Marléne Baumeister$^{*,1,3}$,\\
	 Nicole Krämer$^{3,4}$, Andreas Groll$^{1}$,\\
	  Markus Pauly$^{1,3}$, Magdalena Wischnewski$^{3}$}

\clearpairofpagestyles
\ihead{\headmark}
\automark{section}
\ohead{Trust in Regularisation}
\DeclareMathOperator{\prob}{P}

\definecolor{bblue}{HTML}{28539E}
\definecolor{turquoise}{HTML}{47C2D6}
\definecolor{pink}{HTML}{E39AB7}
\definecolor{lgreen}{HTML}{94e017 }
\definecolor{orange}{HTML}{e88909}

\begin{document}

\thispagestyle{empty}
\maketitle

\renewcommand*{\thefootnote}{\fnsymbol{footnote}}
\footnotetext[1]{These authors contributed equally to this work.}
\footnotetext[2]{Corresponding author: Konstantin Emil Thiel, \textsf{e-mail: thiel@statistik.tu-dortmund.de}}
\renewcommand*{\thefootnote}{\arabic{footnote}}
\footnotetext[1]{Department of Statistics, TU Dortmund University, Germany}
\footnotetext[2]{Core Facility Biostatistics, Paracelsus Medical University Salzburg, Austria}
\footnotetext[3]{Research Center Trustworthy Data Science and Security, UA Ruhr, Germany}
\footnotetext[4]{Department of Social Psychology: Media and Communication, University of Duisburg-Essen, Duisburg, Germany}

\section*{Abstract}

Statistical practice does not automatically follow methodological innovation.
Regularization methods, widely advocated to reduce overfitting and stabilize inference, are readily available in modern software, but are not consistently used by data analysts.
We investigate this implementation gap in a large-scale empirical study of trust in, and acceptance of, regularization techniques, based on $N = 606$ data analysts.
Drawing on measurement frameworks from technology acceptance research, we survey practitioners and embed a randomized experiment to test whether written recommendation of regularization methods increases trust or intended use.
We find no evidence of such an effect.
Instead, adoption intentions are strongly associated with analysts' perceptions of ease of implementation and practical benefit, such as improved bias control or interpretability.
Perceived social norms also emerge as a central driver. 
These results indicate that uptake of statistical methodology depends less on formal recommendations than on usability, perceived utility, and community practice.

\textbf{Keywords:} Technology Acceptance Model, Regularization, \textsc{lasso}, Survey Data, Trust.

\section[Introduction]{Introduction and Motivation}\label{sec:intro}
Statistical research often aims to improve existing methods in terms of their applicability.
Accordingly, we evaluate the performance of new methods with statistical indicators such as level $\alpha$ control, power, and the strength and restrictiveness of the underlying mathematical assumptions.
In addition, a central goal for statisticians is that methodological innovations offering improvements over established techniques are later used by practitioners.
Lynne Billard, former president of the American Statistical Association, pointed out as early as 1997: ``Statistics as a discipline cannot exist by itself.'' \parencite{billard_voyage_1997}.
This implies that the properties and advantages of new statistical methods must be explained to an applying audience.
The effort required to communicate methodological advances effectively has been emphasized from various perspectives just as often as the implementation of this ideal has been criticized.
\textcite{heinze_phases_2023} identify structural problems in the biostatistics community that prevent innovations in statistical methodology from diffusing beyond the methodological community.
They highlight the importance of simulation studies to assess the reliability of new statistical methods and advocate for papers that address applied audiences as well as accessible software implementations.
\textcite{sharpe_why_2013} argues that a persistent gap separates quantitative methodologists, who develop new techniques, from applied researchers, who might benefit from them and highlights that the adoption of new analytical tools is not simply a matter of technical merit or accessibility, but also of communication and social translation.
Both conclude that statistical practitioners do not use improved statistical methods to the extent that would be desirable from a methodological perspective.
But even when statistical practitioners have been adequately addressed, other aspects might prevent them from using new statistical methods.
Potential barriers include low trust in new methods, limited statistical experience, an unsupportive environment that does not encourage engagement with new methods, and the perception that learning new methods is too time-consuming.

This is particularly evident in the context of \textit{regularization methods}. 
This term describes statistical methods that control the complexity of a statistical model to prevent overfitting \parencite{friedrich_regularization_2023} and thereby improves generalizability, enabling meaningful inference and predictions. 
Some examples of regularization methods include penalization, e.g.\ in the form of Least Absolute Shrinkage and Selection Operator \parencite[\textsc{lasso},][]{tibshirani_regression_1996}, ridge regression \parencite{hoerl_ridge_1970}, elastic net \parencite{zou_regularization_2005}, or the use of Bayesian priori distributions \parencite{berger_statistical_1985, park_bayesian_2008}.
Other methods of regularization are early stopping \parencite{friedman_special_2000, friedman_greedy_2001}, e.g.\ in statistical boosting, and ensembling, e.g.\ in the form of bagging \parencite{breiman_bagging_1996} and model averaging \parencite{claeskens_model_2008}.
The multitude of methods and the fact that regularization is a central topic in standard statistics textbooks \parencite[e.g.][]{hastie_elements_2009} already suggests that regularization methods are widely discussed and well-understood in the statistics community.
Already in 2015, \textcite{mcneish_using_2015} reviews modern penalization techniques, highlights their popularity in the statistics community and discusses the underutilization in behavioural science research.
He attributes this to a combination of limited awareness, perceived complexity, and the lack of accessible guidance and, in line with \textcite{heinze_phases_2023}, implementation examples for applied researchers.
Nearly a decade later, \textcite{friedrich_regularization_2023} examined the use of methods of regularization in clinical applications through a systematic literature review. 
They found that the majority of the reviewed medical studies did not use methods of regularization in their statistical analysis.
Further, from a methodological point of view, they state that the use of regularization methods would have been beneficial in several reviewed studies and that software implementations are available in \textsf{R} \parencite{r_core_team_r_2025}, \textsf{SAS} \parencite{sas_institute_inc_sas_2025}, and \textsf{Stata} \parencite{statacorp_llc_stata_2025} for many regularization approaches that could have been used.
\textcite[p.437]{friedrich_regularization_2023} conclude that ``the only downside of the regularization approaches is increased complexity in the conduct of the analyses which can pose challenges in terms of computational resources and expertise on the side of the data analyst''.  
There is also additional work that points out the methodological importance and the simultaneous non-use of regularization for specific application fields \parencite{hoffmann_decisions_2024}.
In summary, the observation of statisticians as well as applied researchers is that regularization is not widely used by statistical practitioners despite its availability as a statistical method.

In this manuscript, we therefore investigate systematically why statistical practitioners often do not adopt regularization methods in practice.
To this end, we have conducted a survey, examining reasons for using or not using regularization, e.g. trust and vigilance \parencite{lewicki_trust_1998, wischnewski_development_2025}, experience, perception of regularization in the own social environment, understanding of regularization as a statistical method, and the expected performance and effort of using regularization in statistical analyses.
We thereby draw on the Unified Theory of Acceptance and Use of Technology \parencite[\textsc{utaut},][]{venkatesh_user_2003} as the underlying psychological concept.
Background on this framework can be found in Section \ref{sec:theory}.
The data analysis is presented in Section \ref{sec:methodsanalysis}.
There, the sample and the experimental design are described (Section~\ref{sec:design}) and the measures and their corresponding items are given (Section~\ref{sec:measures}).
We then analyse the data via a correlation analysis between the psychological constructs and the intention to use regularization in the participant's own analyses (Section \ref{sec:corranalysis}).
The investigation, of whether recommendations by peers, experts or journals influence intention to use regularizations and trust in regularization is presented in Section \ref{sec:facanalysis}.
As an exploratory analysis, we fit a regularized ordinal regression model that provides further insights into how attitudes of statistical practitioners towards regularization influence their intention to use it (Section~\ref{sec:explorative}). 
The manuscript closes with a discussion and an outlook in Section \ref{sec:end}.

\section{Theoretical Background}\label{sec:theory}

In this section, we introduce psychological constructs related to the intention to use regularization methods, which form the theoretical basis for the preregistered hypotheses presented in Sections~\ref{sec:hypotheses} and~\ref{sec:hypotheses_rec}~\parencite{wischnewski_trust_2024}.

\subsection{Psychological Perspectives towards Adaption to Methodological Advances}\label{sec:utaut}

Understanding why certain ideas spread successfully whereas others do not has long attracted scholar attention.
Influential works such as \textcite{rogers_diffusion_2003}, \textcite{granovetter_threshold_1978}, and \textcite{may_implementing_2009} describe innovation uptake as a process shaped by interactions among individuals, social networks, and organizational contexts.
To better understand the individual level, cognitive and affective mechanisms towards innovations must be investigated.
In a statistical context, \textcite{sharpe_why_2013} and \textcite{mcneish_using_2015} observe that practitioners are discouraged from using certain statistical methods. 
They point to key factors such as researchers' perceptions that advanced methods are ``beyond their statistical comfort zone'' \parencite[p.575]{sharpe_why_2013}.

To conceptualize these aspects more systematically, we draw on established models from technology acceptance research.
The Technology Acceptance Model \parencite[\textsc{tam};][]{davis_user_1989} was originally developed to explain computer software adoption and later applied to big data analytics \parencite{al-ateeq_big_2022}, educational technologies \parencite{schorr_technology_2023}, and statistical software use \parencite{wallace_adoption_2014}.
\textcite{davis_user_1989} propose that users' behavioural intention (\textsc{bi}) is the main predictor of later technology use, where \textsc{bi} itself is predicted by additional psychological constructs such as perceived ease of use.

Building upon this foundation, the Unified Theory of Acceptance and Use of Technology \parencite[\textsc{utaut};][]{venkatesh_user_2003} uses the following constructs to model an individual's \textsc{bi}: \emph{performance expectancy} (\textsc{pe}) reflects beliefs about how using an innovation enhances analytical outcomes; \emph{effort expectancy} (\textsc{ee}) captures perceived ease of learning and applying a method; \emph{social influence} (\textsc{si}) refers to perceived endorsement or pressure from peers or supervisors; and \emph{experience} (\textsc{ex}) moderates these relationships.
In addition to these core \textsc{utaut} constructs, \emph{attitude toward using technology} (\textsc{at}) was also considered in the development of \textsc{utaut}, and we therefore include it as a related factor capturing a person's overall orientation toward methodological innovations.
While \textsc{utaut} was developed outside statistics, its constructs can be transferred to the present context.
For instance, expectations that regularization improves prediction accuracy or model stability can be regarded as \textsc{pe}, while high computational or conceptual demands may discourage use, an aspect reflected in \textsc{ee}.
Together, these constructs provide a framework for analysing the uptake of regularization methods in light of perceived benefits, costs, social context, and prior experience.

\subsection{Trust, Vigilance, and System Understanding}\label{sec:trust}

Beyond the variables discussed in \textsc{tam} and \textsc{utaut}, following previous work \parencite{wanner_effect_2022, vorm_integrating_2022}, we also consider trust and users' understanding of regularizations to explain the (missing) uptake.
We assume that users may hesitate to adopt new statistical techniques not only because of perceived complexity or lack of training, but also due to uncertainty about the reliability, validity, or proper application of these methods.
In line with this, \textcite{heinze_phases_2023} speak about \textit{trustworthiness} as an aim in the development of statistical methods.
Empirical studies have further shown that trust in the technology itself, in its developers, or in its endorsers can significantly influence intentions \parencite{mcknight_developing_2002, gefen_trust_2003, pavlou_consumer_2003}.
In the context of regularizations, trust may therefore shape whether researchers experiment with, rely on, and integrate these methods into their workflow.

While early works conceptualize trust as a unidimensional continuum from trust to distrust~\parencite{mayer_integrative_1995}, others argue that trust and distrust are distinct and potentially co-occurring constructs \parencite{lewicki_trust_1998}.
In the context of AI, this co-occurrence has been linked to ``black-box'' systems: users may trust system performance while simultaneously distrusting its limited intelligibility.
Supporting this view, \textcite{wischnewski_development_2025} identify vigilance -- a form of watchfulness -- as a subdimension of trust in AI.
In line with this literature, we conceptualize vigilance as a predictor of regularization uptake, with higher vigilance associated with reduced behavioural intention.

In addition to trust, we consider \emph{system understanding}, that is, the degree to which users comprehend the internal logic and mechanics of a method.
Applied to regularization, users who cannot form a coherent mental model of what, for example, a \textsc{lasso} penalty is doing to their coefficients may be reluctant to trust or rely on (see more on this in the next section) a process whose outcomes feel opaque or arbitrary.

We treat system understanding as conceptually distinct from effort expectancy.
Whereas expectancy concerns the perceived ease of learning and applying a method, system understanding reflects whether users feel they know why the method behaves as it does.
A user may be able to run a regularized regression in software, yet still lack a principled understanding of how the regularization parameter governs the bias--variance tradeoff.
This distinction may help explain why some practitioners recognize the value of regularization but still do not use it in practice.
Moreover, system understanding connects to broader work on algorithm transparency in human--\textsc{ai} interaction, where black-box systems are often associated with lower trust and lower adoption, regardless of predictive performance \parencite{langer_what_2021, mahmud_what_2022}.

\subsection{Hypotheses w.r.t.\ Correlation between Psychological Constructs}\label{sec:hypotheses}
To the best of our knowledge, our work is the first to apply psychological constructs from technology acceptance research to investigate data analysts and statisticians' attitudes towards certain statistical methods.
When we designed the survey, there were open questions on the relation between the constructs, and therefore, one of our central goals was a comprehensive correlation analysis. 
This is reflected in the following preregistered hypotheses \parencite[cf.][]{wischnewski_trust_2024}:

\begin{enumerate}
	\item[\textbf{C1}]
	\begin{enumerate}
		\item Performance expectancy is correlated with intention to use.
    \item Performance expectancy is positively correlated with intention to use.
	\end{enumerate}
	\item[\textbf{C2}]
	\begin{enumerate}
		\item Effort expectancy is correlated with intention to use.
    \item Effort expectancy is positively correlated with intention to use.
	\end{enumerate}
	\item[\textbf{C3}]
	\begin{enumerate}
		\item Social influence is correlated with intention to use.
    \item Social influence is positively correlated with intention to use.
	\end{enumerate}
\item[\textbf{C4}]
	\begin{enumerate}
		\item System understanding is correlated with intention to use.
    \item System understanding is positively correlated with intention to use.
	\end{enumerate}
\item[\textbf{C5}]
	\begin{enumerate}
		\item Vigilance is correlated with intention to use.
    \item Vigilance is negatively correlated with intention to use.
	\end{enumerate}
\item[\textbf{C6}]
	\begin{enumerate}
		\item Trust is correlated with intention to use.
    \item Trust is positively correlated with intention to use.
	\end{enumerate}
\end{enumerate}

Note that we hypothesized consistently positive correlations, with the exception of vigilance.
We will investigate these hypotheses in Section~\ref{sec:corranalysis}.

\subsection{The Effect of Recommendation by Peers, Experts and Journals}\label{sec:recommendation}
Much of our cognition and decision-making is affected by our social environment. 
Prior work suggests that adoption is not solely determined by a method's technical merits or perceived ease of use, but also by who communicates or advocates for it \parencite{venkatesh_user_2003}. 
We therefore examine the role of \emph{recommendations} to understand how endorsements affect the adoption of regularization methods.
The statistical community has also recognized that communication about methods in general, and recommendations in particular, shape their uptake. 
The works of \textcite{mcneish_using_2015} and \textcite{friedrich_regularization_2023} can be understood as recommendations for applied researchers to use regularization methods in their statistical analyses.
Apart from that, \textcite{heinze_phases_2023} encourage the biostatistical community to write tutorial and review papers to provide recommendations for new statistical methods.

We chose to consider three types of recommendations that are well-founded by different aspects of psychological theory.
In addition to experts, also journals provide recommendations, e.g. by specifying expected methodological standards (e.g. in the context of open science goals).
Both journals and experts may represent epistemic authorities \parencite{jager_epistemic_2025} for users of regularizations and may therefore affect their decisions.
Epistemic authority refers to the tendency to trust the statements of others because of their expertise \parencite{mccraw_nature_2015}. 
For example, we commonly trust a medical doctor’s explanation of a diagnosis because we recognize their professional training and expertise in the relevant field. 
Within academia, we suggest that journals and experts are typically granted similar forms of epistemic authority. 

In summary, we hypothesize that recommendations by certain authorities, for instance, advice from a statistical expert, increase the use of regularizations in the scientific community. 
Likewise, we suggest that instructions to use regularization methods in journal guidelines provide an effective way to overcome their underutilization.
In addition, we also examined peer recommendations. We thereby understand peer recommendations as a social norm \parencite{cialdini_focus_1991} that may affect actual usage. 
A social norm refers to a shared expectation within a group about how individuals should think or behave in a given situation. 
Social norms function as informal rules that guide behaviour by signalling what is considered typical or appropriate among others. 
We propose that, when researchers observe that colleagues frequently use or recommend a particular method, they may perceive its use as the accepted or appropriate approach, encouraging them to adopt the method in order to align with disciplinary standards, gain peer approval, or signal methodological competence. 
Consequently, peer recommendations as a social norm may contribute to the diffusion and actual use of specific statistical methods.

\subsection{Hypotheses w.r.t. Recommendation}\label{sec:hypotheses_rec}
In addition to the hypotheses that refer to the \textsc{utaut}, we consider hypotheses that reflect the influence of recommendations to use regularization methods.
The following hypotheses were stated in the preregistration \parencite{wischnewski_trust_2024}:
\begin{enumerate}
	\item[\textbf{R1}] People trust methods of regularization differently, when they are recommended by peers, by experts, by journals or if it is not recommended at all.
	\item[\textbf{R2}] People are vigilant about methods of regularization in different ways, when they are recommended by peers, by experts, by journals or if it is not recommended at all.
	\item[\textbf{R3}] People's intention to use methods of regularization is different, when they are recommended by peers, by experts, by journals or if it is not recommended at all.
\end{enumerate}
In addition to these global hypotheses, we are interested in pairwise local hypothesis, comparing each recommendation group with the control group \textit{no recommendation}. 
In particular, we test whether behavioural intention and trust differ between each recommendation type and the absence of any recommendation. 

\section{Methods and Statistical Analysis}\label{sec:methodsanalysis}
We conducted a survey to address our preregistered hypotheses \parencite[cf.][]{wischnewski_trust_2024}.
The full dataset as well as source code of our data analysis can be found in the Supplement~\parencite{thiel_supplementary_2026}.

\subsection{Sample and Experimental Design}\label{sec:design}

We recruited participants via the online survey platform Prolific \parencite{prolific_easily_2025}.
For sample size planning, the formula for pilot studies by \textcite{viechtbauer_simple_2015} \parencite[see preregistration][]{wischnewski_trust_2024} was used.
Our survey included two questions that served as attention checks and we excluded 52 participants who did not pass these checks.
Moreover, each participant had to answer the screening question \textit{In your job, do you analyse data at least once a month?} with \textit{yes}. 
This screening question reflects our particular interest in a broader sample that allows participants with different levels of expertise to avoid bottom or ceiling effects. 
People with a degree in statistics would be too familiar with regularizations, making our intervention not  meaningful.
40 participants were excluded by failing the screening question.
Furthermore, we excluded 108 participants who quit the survey early and therefore did not respond to all items explained in Section~\ref{sec:measures}.
One additional participant was excluded due to a typo in the age field. 
Finally, $N = 606$ ($306$ male, $292$ female, $4$ non-binary, $4$ preferred not to say) participants were included in our analysis.
To test our hypotheses, we developed a four group between-subjects design to which the participants were randomly assigned.
The groups were \textit{control} ($n_1=156$), \textit{expert} ($n_2=145$), \textit{journal} ($n_3=158$), and \textit{peer} ($n_4=147$).
Participant age ranged from 18 to 87 (median $m=30$, interquartile range $r=13$).
The participants reported their highest degree received: one had completed middle school, $81$ high school, $329$ held a Bachelor's degree, $148$ a Master's degree, $29$ a PhD, and $18$ reported none of the listed categories.
They indicated the following as their fields of study: \textit{Data Science/Statistics} ($212$), \textit{Economics and Business} ($77$), \textit{Computer Science} ($68$), \textit{Engineering} ($50$), \textit{Psychology} ($38$), \textit{Medicine} ($31$), \textit{Biology/Geography/Agriculture} ($16$), \textit{Mathematics} ($7$), and \textit{Other} ($107$).
In terms of geographical residency, our participants can be summarized as follows: \emph{South Africa} ($213$), \emph{United Kingdom} ($153$), \emph{Europe} ($153$), \emph{North America} ($60$), \emph{Kenya} ($10$), \emph{other countries} ($11$), and 5 participants chose not to disclose.
This distribution likely reflects that the Prolific survey was released at 11am Central European Time and targeted English-speaking individuals.
All three countries, South Africa, United Kingdom, and Kenya, are located in similar time zones.

We introduced participants to regularization methods using a short explanation and two examples inspired by the notion of regularization discussed in \textcite{friedrich_regularization_2023}.
The full examples can be found in the Supplementary Material \parencite{thiel_supplementary_2026}.
Afterwards, participants read a recommendation statement in which the source of the recommendation depended on their assigned group: 
\textit{In sum, regularizations are a useful technique to improve your statistical analyses.
This is why
\begin{enumerate}
	\item \text{(Peer Recommendation)} researchers from various disciplines increasingly recommend using regularizations. They are among the fastest-growing ways of analysis of the last 20 years.
	\item \text{(Journal Recommendation)} leading journals strongly recommend the implementation of regularizations. Some journals already consider to desk-reject manuscripts that do not use regularizations even though this would be appropriate.
	\item \text{(Expert Recommendation)} leading statistical experts strongly recommend the implementation of regularizations. For example, \textcite{friedrich_regularization_2023} give convincing evidence on why these analyses indeed lead to more valid results.
\end{enumerate}}
In the control group, no recommendation statement was presented.
Subsequently, participants responded to statements about their attitudes toward regularization as explained in Section~\ref{sec:measures}.
At the end of the survey, all participants were asked about the type of recommendation which they were assigned to (manipulation check), afterwards, they were debriefed.

\subsection{Measures}\label{sec:measures}

Our survey aimed to measure the psychological constructs described in Section~\ref{sec:theory} using 5-point Likert items, ranging from 1 (\emph{strongly disagree}) to 5 (\emph{strongly agree}).
Table~\ref{tab:items} lists all constructs together with their associated items.
Note that the number of items per construct varies, with six items used for performance expectancy (\textsc{pe}) and system understanding (\textsc{su}), whereas behavioural intention (\textsc{bi}) was measured with a single item.
The individual items were aggregated into mean scores such that one score per construct was obtained.
Aggregation into mean scores was justified by internal consistency checks using Cronbach's $\alpha$~\parencite{cronbach_1951}.
We observed high $\alpha$ values across all constructs, ranging from $\alpha = 0.77$ for \textsc{pe} to $\alpha = 0.88$ for experience (\textsc{ex}), effort expectancy (\textsc{ee}), and attitude towards technology (\textsc{at}), implying that, for each construct, the individual items measure the same underlying psychological construct.
For social influence (\textsc{si}), however, a satisfying value of $\alpha = 0.82$ could only be reached after excluding a third item (\emph{Using regularizations would not be accepted at my work. [reversed]}), which turned out to be uncorrelated with the remaining two \textsc{si} items listed in Table~\ref{tab:items}.

\begin{figure}[ht]
	\centering
	\includegraphics[width=0.8\textwidth]{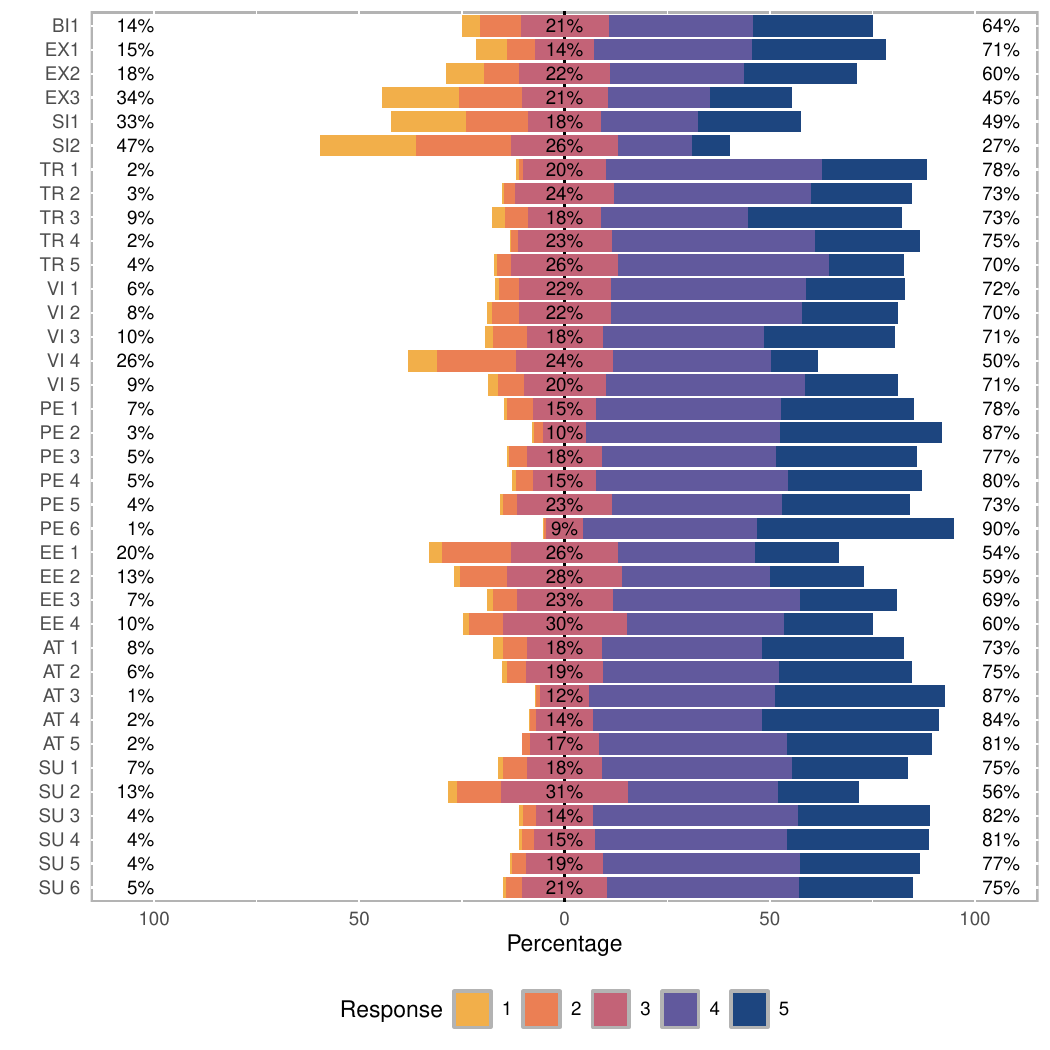}
	\caption{Distribution of the 5-point Likert items in the order given in Table~\ref{tab:items}. The percentage on the left hand side/middle/right hand side shows the (combined) share of response categories 1 and 2/3/4 and 5.}
	\label{fig:items}
\end{figure}

Figure~\ref{fig:items} shows the distribution of the individual Likert items.
We observe a tendency towards positive responses, typical for survey data~\parencite{bogner_2016}, in particular for the items of \textsc{tr}, \textsc{vi}, and \textsc{pe}.
For \textsc{ee}, \textsc{ex}, and \textsc{bi}, this tendency is less pronounced, and for \textsc{si}, the items follow a balanced distribution.
In our subsequent analyses, we treat all mean scores constructed from two or more items as pseudo-metric variables.
Since \textsc{bi} consists of only one item, we treat \textsc{bi} as an ordinal variable.
Importantly, due to the moderate balance of the data, all categories 1 -- 5 of \textsc{bi} occur in the data, with a minimum number of 26 observations in category 1.
This allows us to use modelling tools for ordinal data, such as the proportional odds logistic regression.

\begin{longtable}{p{0.21\textwidth} p{0.74\textwidth}}
	\caption{\label{tab:items}Constructs with the surveyed 5-point Likert items.}\\

	\toprule
	\textbf{Construct} & \textbf{Items} \\ 
	\midrule

	(\textsc{bi}) Behavioural Intention %
	& 1: I intend to use regularizations for my next projects. \\[3pt]
	\midrule

	\multirow{3}{=}{(\textsc{ex}) Experience} %
	& 1: I have heard about regularizations as a statistical method. \\* 
	& 2: I have read about regularizations as a statistical method. \\*
	& 3: I have applied regularizations as a statistical method. \\[3pt]
	\midrule
	
	\multirow{2}{=}{(\textsc{si}) Social Influence} %
	& 1: I know people who use regularizations. \\* 
	& 2: People around me expect me to use regularizations. \\[3pt]
	\midrule

	\multirow{5}{=}{(\textsc{tr}) Trust} %
	& 1: I believe in (the results of) regularizations. \\* 
	& 2: I trust (the results) of regularizations. \\* 
	& 3: I distrust (the results of) regularizations. [reversed] \\* 
	& 4: Regularizations are developed to be trustworthy. \\* 
	& 5: I can depend on (the results of) regularizations. \\[3pt]
	\midrule
	
	\multirow{5}{=}{(\textsc{vi}) Vigilance} %
	& 1: I am vigilant about the interpretability of statistical methods with regularization. \\* 
	& 2: I am alert of the interpretability of statistical methods with regularization. \\* 
	& 3: I am on the lookout for errors when using regularizations. \\* 
	& 4: I am on my toes when statistical methods use regularizations. \\*
	& 5: I am careful when statistical analyses use regularizations. \\[3pt]
	\midrule
	
	\multirow{6}{=}{(\textsc{pe}) Performance Expectancy} %
	& 1: Regularizations would help me to do statistical analyses faster. \\* 
	& 2: Using regularizations would improve my statistical analyses. \\* 
	& 3: Regularizations would make my statistical analyses easier to interpret. \\* 
	& 4: I think using regularizations would be useful for my empirical work. \\*
	& 5: Regularizations can reduce statistical bias in my analyses. \\*
	& 6: Regularizations can be applied in various scenarios.\\[3pt]
	\midrule
	
	\multirow{4}{=}{(\textsc{ee}) Effort Expectancy} %
	& 1: Applying regularizations would be easy for me. \\* 
	& 2: It would be easy for me to apply regularizations in a way that would benefit me. \\* 
	& 3: Applying regularizations would be understandable to me. \\* 
	& 4: Learning how to use regularizations would be easy for me. \\[3pt]
	\midrule
	
	\multirow{5}{=}{(\textsc{at}) Attitude towards Technology} %
	& 1: Applying regularizations could make my work more sophisticated. \\* 
	& 2: In principle, I would like to work with regularizations. \\* 
	& 3: Using regularizations could be a good idea. \\* 
	& 4: Using regularizations could be a wise decision. \\*
	& 5: Regularizations can provide unique statistical decisions beyond common procedures.\\[3pt]
	\midrule

	\multirow{6}{=}{(\textsc{su}) System Understanding} %
	& 1: I understand regularization as a statistical method. \\ 
	& 2: I understand the limitations of regularizations. \\ 
	& 3: I understand what regularization can add to a statistical analysis. \\ 
	& 4: I understand how regularization contributes to statistical decision making. \\		
	& 5: I understand how regularization contributes to statistical modelling. \\
	& 6: I understand what regularization could add to statistical variable selection.\\[3pt]
	\bottomrule
\end{longtable}

\subsection{Correlation Analysis}\label{sec:corranalysis}

\begin{figure}[t]
	\centering
	\includegraphics[width=0.55\textwidth]{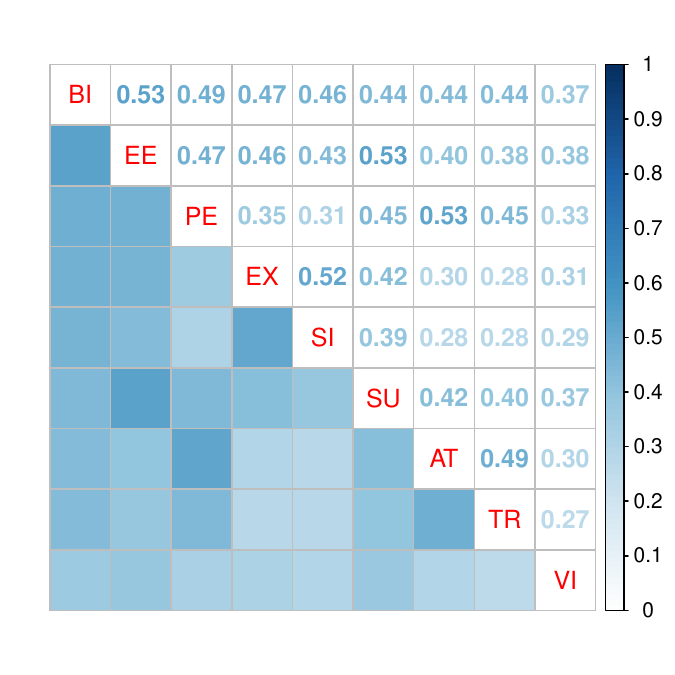}
	\caption{Empirical Kendall's~$\tau$ correlation matrix of the investigated constructs. Constructs are sorted in descending order of their correlation with \textsc{bi}.}
	\label{fig:kendall-correlation}
\end{figure}

In our study preregistration \parencite{wischnewski_trust_2024}, we planned to investigate the association between behavioural intention (\textsc{bi}) and other constructs via correlation analysis.
Since \textsc{bi} is a discrete ordinal variable, we use Kendall's~$\tau$ and plot the resulting correlation matrix (without diagonal entries) in Figure~\ref{fig:kendall-correlation}. The matrix shows moderately positive correlations across all constructs.
That is, individuals who score high on one construct tend to score high on others, which suggests consistent attitudes toward regularization methods.
To investigate the correlation structure in detail and to address the preregistered analysis (cf.~Section~\ref{sec:hypotheses}), we formally test the following hypotheses:
\begin{align*}
\mathcal{H}_0: \tau = 0 && \text{vs.} && \mathcal{H}_1: \tau \neq 0,
\end{align*}
as well as the one-sided counterpart 
\begin{align*}
\mathcal{H}_0: \tau \leq 0 && \text{vs.} && \mathcal{H}_1: \tau > 0,
\end{align*}
where $\tau$ denotes one of $36$ possible Kendall correlation coefficients between pairs of constructs from Figure~\ref{fig:kendall-correlation}.
When testing for significance at $\alpha = 0.025$ while controlling for multiplicity using 
a Bonferroni correction, we obtain that all pairwise correlations are significantly positive. 
These results are based on asymptotic tests for Kendall's~$\tau$. 
As a robustness check, we additionally tested for linear correlations among the quasi-metric constructs using permutation tests based on Pearson's $\rho$ (Bonferroni adjusted), obtaining consistent results across these two correlation measures \parencite[cf.][]{thiel_supplementary_2026}.

Among all constructs, \textsc{bi} shows some of the strongest associations with other constructs, with $\hat \tau (\textsc{bi}, \textsc{ee}) = 0.53$ representing the largest observed correlation.
Individuals with high effort expectancy (\textsc{ee}) consider applying regularization methods to be easy, which is in line with the high correlation with \textsc{bi}.
At first sight, it appears less intuitive why vigilance (\textsc{vi}) is positively correlated with \textsc{bi}, as the term might suggests the opposite (i.e., a negative correlation).
In fact, we even hypothesized a negative correlation in our study preregistration (cf.~Section~\ref{sec:hypotheses}).
However, reviewing the specific \textsc{vi} items from Table~\ref{tab:items}, we interpret vigilance as reflecting critical attentiveness towards regularization methods.
Such individuals may not refrain from using regularization methods, but instead may be particularly careful when applying them, which provides a plausible explanation for observed positive correlation.
In summary, hypotheses \textbf{C1} -- \textbf{C4} and \textbf{C6} are supported, while we found a reversely directed correlation for \textbf{C5}.

\begin{figure}[th]
	\centering
	\includegraphics[width=0.8\textwidth]{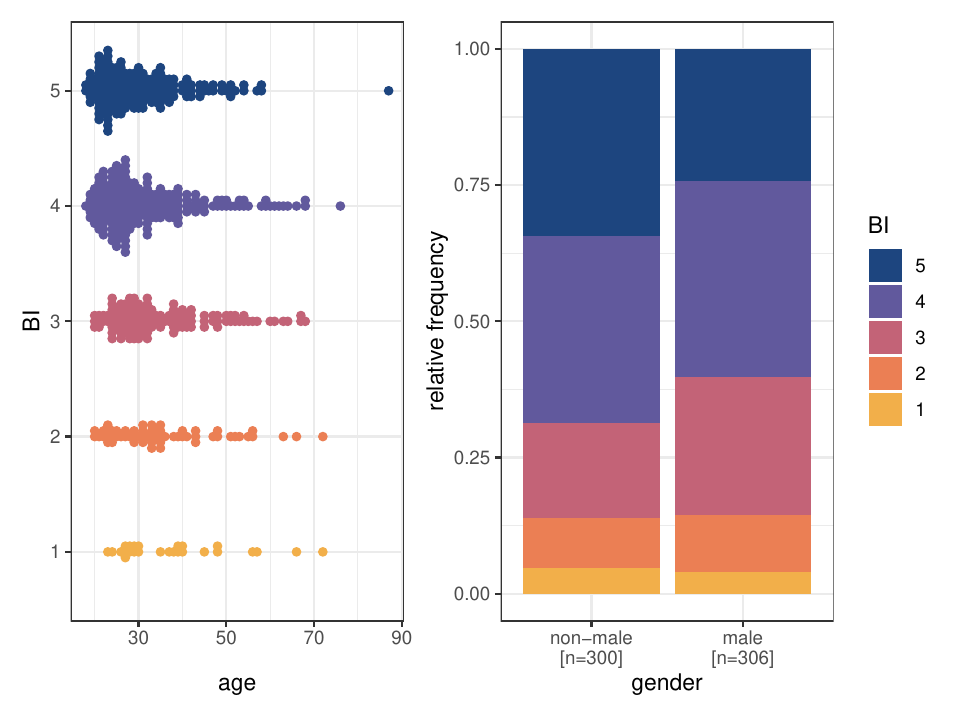}
	\caption{Joint distribution of demographic variables with \textsc{bi}.}
	\label{fig:demographics-bi}
\end{figure}

Figure~\ref{fig:demographics-bi} displays the association of demographic variables with \textsc{bi}.
We note that \textsc{age} and \textsc{bi} are negatively correlated, with Kendall's~$\tau$ yielding $\hat \tau = -0.16$.
This suggests that older individuals may be more hesitant to adopt regularization methods. 
Moreover, Figure~\ref{fig:demographics-bi} suggests a gender effect, where individuals identifying non-male tend to respond more frequently in the  higher \textsc{bi} categories 4 and 5, while individuals who identify as male respond more frequently in the central category 3.
This pattern thus suggests that non-male individuals might be more open to regularization uptake than male.
In general, the notable associations between \textsc{bi} and other variables suggest that \textsc{bi} may be modelled as a dependent variable of psychological constructs and demographic variables.
We explore this in an explorative analysis in Section~\ref{sec:explorative}.

\subsection{Factorial Analysis}\label{sec:facanalysis}
\begin{figure}[t]
	\includegraphics[scale=0.48]{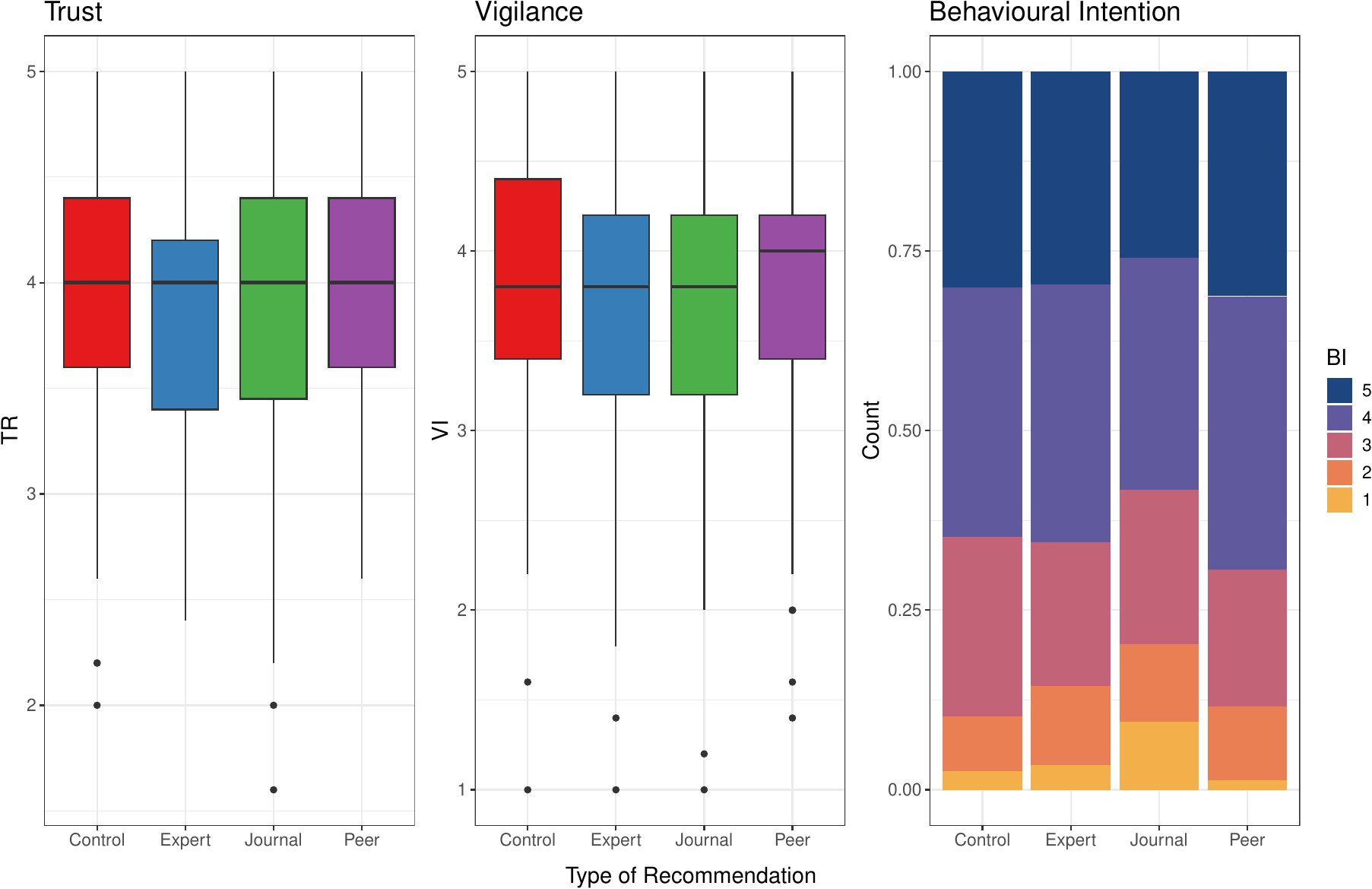}
	\caption{Scores for the constructs trust (\textsc{tr}), vigilance (\textsc{vi}), and behavioural intention (\textsc{bi}) per recommendation groups \textit{Control, Expert, Journal, and Peer}.}
	\label{fig:descr_factorial}
\end{figure}

As described in Section \ref{sec:hypotheses_rec}, we test the three global hypotheses \textbf{R1}, \textbf{R2}, and \textbf{R3} that refer to the constructs \textsc{tr}, \textsc{vi}, and \textsc{bi}.
In Figure~\ref{fig:descr_factorial}, scores of these constructs are plotted across groups.
The figure suggests that the distribution of the constructs \textsc{tr}, \textsc{vi}, and \textsc{bi} are similar across groups.
For \textsc{tr} and \textsc{vi}, the medians are also comparable across groups. 
The construct \textsc{tr} tends to have slightly lower values in the expert group.
For \textsc{vi}, the peer recommendation group seems to be slightly more left-skewed than the others.
For \textsc{bi}, all groups show a tendency toward higher categories which is least pronounced in the journal recommendation group.
We infer whether these differences are statistically significant by testing hypotheses \textbf{R1}, \textbf{R2}, and \textbf{R3}.

Recall that \textsc{bi} consists of only one item (Table~\ref{tab:items}) and has an ordinal scale.
As stated in the preregistration \parencite{wischnewski_trust_2024}, we consider both the global hypotheses \textbf{R1}, \textbf{R2}, and \textbf{R3} and pairwise many-to-one comparisons with the control group.
To handle this and the ordinal scaling of \textsc{bi}, we use nonparametric multiple contrast test procedures based on ranks \parencite{konietschke_rank-based_2012}, implemented in the \textsf{R}-Package \texttt{nparcomp} \parencite{konietschke_nparcomp_2015}.
Let $\alpha=0.05$ be the global significance level. 
For each construct \textsc{tr}, \textsc{vi}, and \textsc{bi}, we test local hypotheses comparing the expert, journal, and peer groups $\{e,j,p\}$ with the control group $c$.
Let $F_\ell^{\textsc{tr}}$, $\ell\in\{c,e,j,p\}$, denote the group distributions of the construct \textsc{tr} with $x^{\textsc{tr}}_\ell\sim F^{\textsc{tr}}_\ell$ and let $z^{\textsc{tr}}$ be distributed according to the \emph{average distribution}, $z^{\textsc{tr}} \sim 1/4\sum_{\ell=1}^4F_\ell^{\textsc{tr}}$.
Then, for group comparisons, we use nonparametric \emph{relative effects}, which can be defined as
$$p^{\textsc{tr}}_{\ell} = \prob\left(z^{\textsc{tr}}<x^{\textsc{tr}}_{\ell}\right) + \frac{1}{2} \cdot \prob\left(z^{\textsc{tr}}=x^{\textsc{tr}}_{\ell}\right),\,\ell\in\{c,e,j,p\},$$
for the construct \textsc{tr}.
If $p^{\textsc{tr}}_{e} = p^{\textsc{tr}}_c$ for example, the group distributions $F^{\textsc{tr}}_e$ and $F^{\textsc{tr}}_c$ of the construct \textsc{tr} are \textit{stochastically comparable} \parencite{brunner_rank-_2018}.
For \textsc{vi} and \textsc{bi}, relative effects $p^{\textsc{vi}}_{\ell}$ and $p^{\textsc{bi}}_{\ell}$ can be defined analogously for every $\ell\in\{c,e,j,p\}$.
Using $p^{\textsc{tr}}_{\ell}$, $\ell\in\{e,j,p\}$, we define the multiple testing problem as 
\begin{align*}
	\mathcal{H}_{0,\ell}^{\textsc{tr}}:\,p^{\textsc{tr}}_{\ell}=p^{\textsc{tr}}_{c}\quad \text{vs.} \quad 
	\mathcal{H}_{1,\ell}^{\textsc{tr}}:\,p^{\textsc{tr}}_{\ell}\not=p^{\textsc{tr}}_{c},
\end{align*}
and use the analogous statements with $p^{\textsc{vi}}_{\ell}$ and $p^{\textsc{bi}}_{\ell}$ for the constructs \textsc{vi} and \textsc{bi}.
The global hypothesis $\mathcal{H}_0^{\textsc{tr}}=\bigcap_{\ell\in\mathcal{R}}\mathcal{H}_{0,\ell}^{\textsc{tr}}$, analogously for \textsc{vi} and \textsc{bi}, is rejected if at least one of the corresponding local hypotheses is rejected. 
Equivalently, the global hypothesis is rejected if at least one of the local $p$-values is smaller than the global level $\alpha$.
Table~\ref{tab:factorialanalysis} shows the results given as local $p$-values as well as the corresponding estimates of the relative effect.

In line with Figure~\ref{fig:descr_factorial}, differences between groups are rather small. 
Only one local hypothesis $\mathcal{H}_{0,e}^{\textsc{tr}}$ is rejected. 
As a consequence, the corresponding global hypothesis $\mathcal{H}_0^{\textsc{tr}}$ is also rejected. 
Overall, the conducted hypothesis tests suggest only minor differences between the groups.
For \textsc{tr}, the expert group differs from the control group, but it differs downward ($\hat{p}^{\textsc{tr}}_{e}=0.46<0.55=\hat{p}^{\textsc{tr}}_{c}$).
In particular, we find no evidence that individuals in any recommendation group exhibit significantly higher trust, vigilance, or intention to use than those in the control group. 
These results suggest that the recommendation manipulation had little to no effect on attitudes towards regularization.
We return to this result in Section~\ref{sec:end}.

\begin{table}[htbp]
	\centering
	\caption{Results of the factorial analysis regarding recommendation. The local $p$-values have to be compared with the global level $\alpha=0.05$. Significances are printed in \textbf{bold}.}
	\label{tab:factorialanalysis}
	\begin{tabular}{lrr}
		
		\toprule
		\textbf{Hypothesis} & \textbf{Estimate} $\hat{p}_{\ell}-\hat{p}_{c}$ & \textbf{Local} $p$-\textbf{Value} \\ 
		\midrule
		$\mathcal{H}_{0,e}^{\textsc{tr}}$ & -0.09	& $\mathbf{0.0260}$ \\
		$\mathcal{H}_{0,j}^{\textsc{tr}}$ & -0.07	& 0.0922 \\
		$\mathcal{H}_{0,p}^{\textsc{tr}}$ & -0.03 & 0.7057 \\
		\midrule
		$\mathcal{H}_{0,e}^{\textsc{vi}}$ & -0.05 & 0.3840 \\
		$\mathcal{H}_{0,j}^{\textsc{vi}}$ & -0.06 & 0.2061 \\
		$\mathcal{H}_{0,p}^{\textsc{vi}}$ & <0.01 & 0.9991 \\
		\midrule
		$\mathcal{H}_{0,e}^{\textsc{bi}}$ & -0.01 & 0.9939 \\
		$\mathcal{H}_{0,j}^{\textsc{bi}}$ & -0.05 & 0.2376 \\
		$\mathcal{H}_{0,p}^{\textsc{bi}}$ & 0.02 & 0.9239 \\
		\bottomrule
	\end{tabular}
\end{table}

\subsection{Explorative Analysis: Predictors of Behavioural Intention}\label{sec:explorative}

While the previous section suggests that a recommendation-based intervention does not explain participant's behavioural intention (\textsc{bi}), we found several variables that may serve as predictors of \textsc{bi} in Section~\ref{sec:corranalysis}.
However, the correlation analysis also revealed dependencies among these predictor candidates, indicating that they should not be used all together in a prediction model without further selection.
Moreover, it remains unclear whether interaction effects between some of the candidates are relevant for predicting \textsc{bi}.
As this is, to the best of our knowledge, the first study on regularization via psychological constructs, no empirical results were available to select a subset of relevant predictors.
Therefore, we will address this question using regularized regression~\parencite{tibshirani_regression_1996} as a variable selection technique.

\subsubsection{Modelling}

Recall that \textsc{bi} takes five ordered categories and is treated as ordinal. 
We use the \emph{cumulative logit model} (\emph{proportional odds logistic regression}) as the robust standard approach for ordinal outcomes~\parencite{inerle_simulation_2026}.
The model assumes an underlying continuous latent variable~\parencite{tutz_regression_2011}, which is linked to the ordinal outcome via thresholds. 
An advantage is that the same set of coefficients is used across categories, while only the intercepts are category-specific, effectively reducing model complexity. 
In our case, the model reads as follows: let $\textsc{bi}^*$ denote the latent version of \textsc{bi}.
Then
$$
\textsc{bi}^* = \boldsymbol{\beta}^\top \mathbf{x} + \varepsilon,
$$
where $\mathbf{x}$ is the vector of covariates and $\varepsilon$ is an error term assumed to follow a logistic distribution.
Moreover, a set of monotonously increasing intercepts $-\infty = \beta_{0, 0} < \beta_{0, 1} < \ldots < \beta_{0, 4} < \beta_{0, 5} = \infty$ links the latent variable back to the observed one:
$$
\textsc{bi} = k \iff  \beta_{0, k - 1} < \textsc{bi}^* \leq \beta_{0, k}.
$$
Finally, we have
\begin{equation}
	\label{eq:polr}
	P(\textsc{bi} \leq k \,|\, \mathbf{x}) = P(\boldsymbol{\beta}^\top \mathbf{x} + \varepsilon \leq \beta_{0, k}) = F(\beta_{0, k} - \boldsymbol{\beta}^\top \mathbf{x}),
\end{equation}
where $F$ is the logistic distribution function.
Note that inversion yields a representation in terms of log-odds:
\begin{equation}
	\label{eq:polr-or}
\log \frac{P(\textsc{bi} > k \,|\, \mathbf{x})}{P(\textsc{bi} \leq k \,|\, \mathbf{x})} = \boldsymbol{\beta}^\top \mathbf{x} - \beta_{0, k},
\end{equation}
that is, we model the odds of \textsc{bi} falling into a higher category given $\mathbf{x}$.

\subsubsection{A two-step LASSO approach}

\begin{figure}[t]
	\centering
	\includegraphics[width=0.7\textwidth]{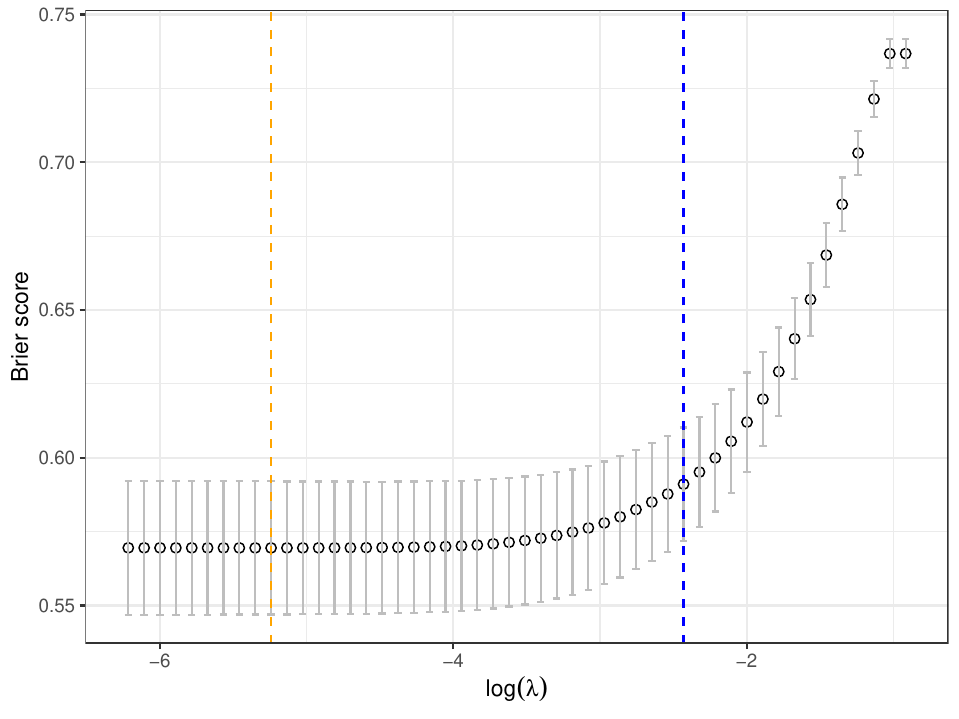}
	\caption{10-fold cross validation Brier score in a cumulative logit model with \textsc{bi} as the outcome variable and the following predictors (main effects only): \textsc{gender}, \textsc{age}, \textsc{ee}, \textsc{su}, \textsc{pe}, \textsc{at}, \textsc{ex}, \textsc{tr}, \textsc{si}, and \textsc{vi}. Grey bars: standard error of the respective cross validation average. Vertical dashed orange line: optimum. Vertical dashed blue line: largest $\lambda$, where the score is within one standard error of the optimum.}
	\label{fig:cv_brier_score_step_1}
\end{figure}

\begin{figure}[t]
	\centering
	\includegraphics[width=0.75\textwidth]{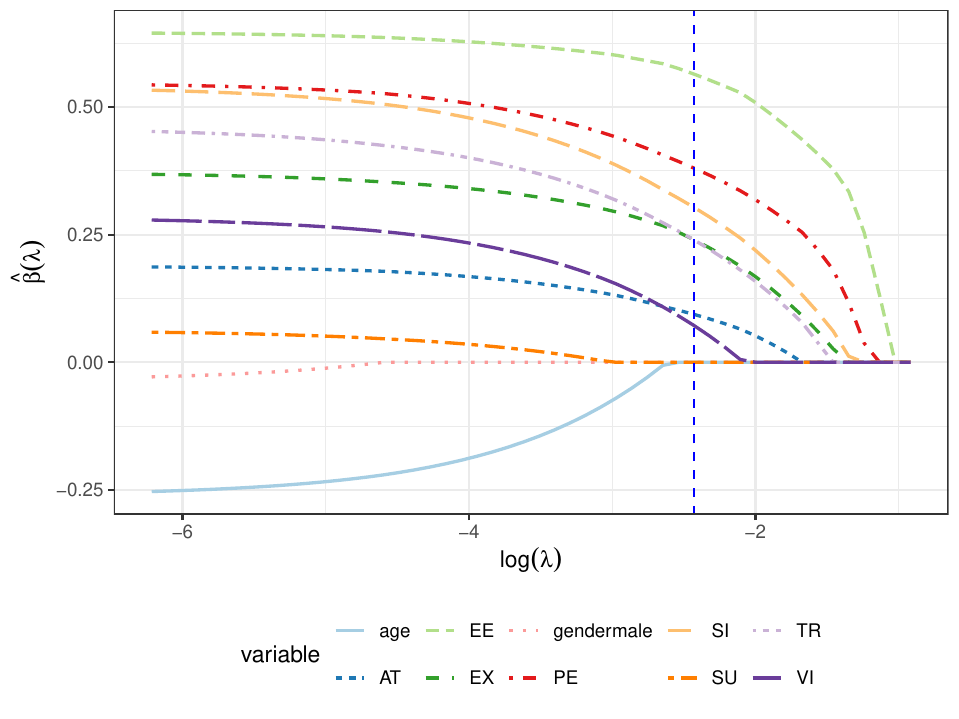}
	\caption{Coefficient paths across \textsc{lasso} penalization in the first step. All variables are scaled to unit variance by dividing through their sample standard deviations. The vertical dashed blue line marks the $\lambda$ value for the 1-SE Brier score from Figure~\ref{fig:cv_brier_score_step_1}.}
	\label{fig:coefficient_path_step_1}
\end{figure}

\begin{figure}[t]
	\centering
	\includegraphics[width=0.7\textwidth]{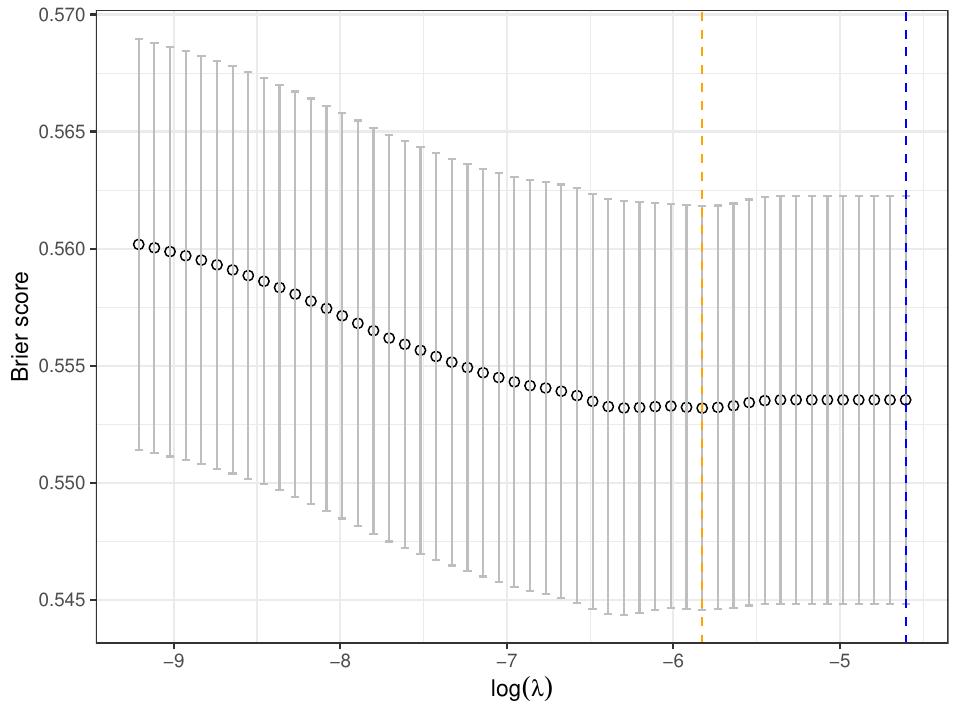}
	\caption{10-fold cross validation Brier score in a cumulative logit model with \textsc{bi} as the outcome variable and predictors \textsc{ee}, \textsc{pe}, \textsc{at}, \textsc{ex}, \textsc{tr}, \textsc{si}, \textsc{vi}, plus all pairwise interactions between these variables. Grey bars: standard error of the respective cross validation average. Vertical dashed orange line: optimum. Vertical dashed blue line: largest $\lambda$, where the score is within one standard error of the optimum.}
	\label{fig:cv_brier_score_step_2}
\end{figure}

\begin{table}[t]
\centering
\caption{Standardized coefficients of the final regularized cumulative model, corresponding to the vertical blue dashed line in Figure~\ref{fig:coefficient_path_step_1}.}
\label{tab:final_regularized_model}
	\begin{tabular}{lr}
		\toprule
		Predictor & Coefficient \\
		\midrule
		\textsc{ee} & 0.565 \\
		\textsc{pe} & 0.380 \\
		\textsc{si} & 0.304 \\
		\textsc{ex} & 0.240 \\
		\textsc{tr} & 0.240 \\
		\textsc{at} & 0.094 \\
		\textsc{vi} & 0.072 \\
		\bottomrule
	\end{tabular}
\end{table}

In a first step, we fit penalized cumulative logit models using \textsc{gender}, \textsc{age}, \textsc{ee}, \textsc{su}, \textsc{pe}, \textsc{at}, \textsc{ex}, \textsc{tr}, \textsc{si}, and \textsc{vi} as predictors.
We include main effects of all candidate predictors identified in the correlation analysis.
We apply \textsc{lasso} penalization using the \texttt{ordinalNet} \textsf{R}-package~\parencite{wurm_regularized_2021}.
Note that \texttt{ordinalNet} assumes an inverse sign convention for the linear predictor. 
We therefore multiply the coefficient estimates obtained from this package by $-1$ to match our notation.
We use 10-fold cross validation to tune the L$_1$-type penalty parameter $\lambda$.
To ensure balanced penalization, all variables are scaled to unit variance.
We aim to identify relevant predictors by selecting a sparse model.
To this end, we apply the \emph{1‑SE rule}: we select the largest $\lambda$ whose cross‑validated Brier score is within one standard error of the minimum~\parencite{hastie_elements_2009}.

After identifying a subset of predictors in this first step, we include their main effects and all corresponding pairwise interactions in a second regularization step.
In this step, we again apply \textsc{lasso} with the 1-SE rule to assess whether interaction terms improve prediction.
To facilitate interpretability, the main effects are excluded from penalization, and therefore, included in this model.

Figure~\ref{fig:cv_brier_score_step_1} reports the cross-validated average Brier score across $\lambda$ values in the first step.
Figure~\ref{fig:coefficient_path_step_1} shows the corresponding coefficient paths $\hat \beta(\lambda)$.
Across the range of $\lambda$, coefficient signs are stable and the paths appear monotonic, suggesting a stable regularization pattern for this fit.
The demographic variables \textsc{age} and \textsc{gendermale} have negative coefficients for small $\lambda$ values but shrink to zero early as $\lambda$ increases.
In contrast, all other constructs have consistently positive coefficients for small $\lambda$, suggesting a positive association with higher categories of \textsc{bi}.

Following the 1-SE rule, we select $\lambda = e^{-2.43}$, which yields a model that retains 7 out of 8 constructs (all except from \textsc{su}) and excludes both demographic variables. 
The fact that most constructs remain suggests that several predictors contribute complementary information.
Among the selected constructs, \textsc{ee}, \textsc{pe}, and \textsc{si} have the largest absolute coefficients.
Since variables are scaled to unit variance, this indicates relatively stronger contributions to the linear predictor for \textsc{bi}.
The corresponding standardized coefficients of the final regularized cumulative model are reported in Table~\ref{tab:final_regularized_model}.
All retained constructs have positive coefficients, indicating that higher values are associated with higher categories of \textsc{bi}.

In the second step, we include all selected main effects (i.e. all excluding \textsc{su}) and their pairwise interactions.
Penalizing the interaction effects only, and following the 1-SE rule, we obtain a model that shrinks all interaction effects to zero, thus resulting in the same model as before in terms of retained variables.
Figure~\ref{fig:cv_brier_score_step_2} shows the corresponding Brier score curve, where the flat part of the curve reflects the model with main effects only.
These results suggest that interaction effects provide little additional value to predict \textsc{bi} beyond main effects.
In Section~\ref{sec:end}, we further interpret these findings.

\section{Discussion}\label{sec:end}
In this work, we have investigated using survey data why statistical practitioners do not use regularization methods despite their availability and widespread discussion in the statistical community.
To this end, we surveyed practitioners about their assessment of regularization methods and their intention to use them.
Thereby, we considered an intentionally broad definition of \textit{statistical practitioners} that includes individuals who conduct statistical analyses on a regular basis.
The considered items refer to constructs from the \textsc{utaut} framework \parencite{venkatesh_user_2003}, as well as trust and vigilance \parencite{wischnewski_development_2025}.
Moreover, participant's attitudes were experimentally manipulated through a recommendation by peers, experts, and journals.

Regarding the preregistered hypotheses we found no consistent evidence that recommendations affect trust, behavioural intention, or vigilance.
Observed differences are small and not systematically in the expected direction. 
Furthermore, we found that all constructs are significantly positively correlated with behavioural intention.
Practically, this means that higher values on these constructs tend to coincide with a stronger intention to use regularizations. 
In contrast to the preregistered hypothesis, the construct vigilance is positively correlated with behavioural intention.
Vigilance can be interpreted as critical attentiveness.
In our context, this suggests that participants with a vigilant attitude are not discouraged from using regularizations.
Overall, the results indicate a high to moderate correlation (measured with Kendall's~$\tau$) between the \textsc{utaut} constructs, trust, vigilance, and behavioural intention.

In addition to the preregistered investigations, we conducted exploratory analyses to identify predictors of behavioural intention.
Using a regularized cumulative logit model for ordinal outcomes, we found that seven of the eight constructs contribute to predicting behavioural intention, with system understanding (\textsc{su}) being the only construct excluded under \textsc{lasso} penalization (cf.\ Table~\ref{tab:final_regularized_model}). 
The strongest predictors for behavioural intention were effort expectancy ($\hat{\beta}^{\textsc{ee}} = .565$), performance expectancy ($\hat{\beta}^{\textsc{pe}} = .380$), and social influence ($\hat{\beta}^{\textsc{si}} = .304$). 
A plausible explanation is that \textsc{su} and \textsc{ee} capture overlapping information, which is in line with their comparatively strong correlation $\hat{\tau}(\textsc{ee},\textsc{su}) = .53$. 
Accordingly, the exclusion of \textsc{su} from the penalized model should not be interpreted as irrelevance, but rather as a lack of additional predictive information beyond other constructs.
These findings suggest that individuals are more likely to intend to use regularization methods when they expect them to be easy to implement (\textsc{ee}), useful (\textsc{pe}), and  experience encouragement from their social environment (\textsc{si}).

The results highlight the importance of usability, perceived utility, and social context for the adoption of statistical methods.
From a statistical perspective, the benefits of regularization methods are well known and easy to apply.
However, our findings indicate that these benefits must also be accessible to practitioners. 
This underscores the importance of tutorial and review papers targeted at applied audiences, characterized as phase IV papers by \textcite{heinze_phases_2023}.
In addition, the importance of effort expectancy in our exploratory analysis, highlights the importance of statistical training and knowledge as also discussed in \textcite{friedrich_is_2022}.
Finally, the importance of social influence indicates that adoption is affected by the research environment which needs to be open for statistical innovations.

As we found no positive effect of direct recommendations on adoption of regularization, it seems that supportive environments and perceived usability matter more. 
In particular, statistical competence and the ability to evaluate new statistical methods appear to be more important than stated beliefs or external recommendations.
Recommendation-based appeals or a \textit{hype} around certain statistical methods appear to have only limited effect on the actual use behaviour.
At the same time, the positive correlation $\hat{\tau}(\textsc{bi},\textsc{tr}) = .44$ of behavioural intention with trust and the positive correlation $\hat{\tau}(\textsc{bi},\textsc{vi}) = .37$ of behavioural intention with vigilance (cf. Figure~\ref{fig:kendall-correlation}) suggest that adoption of regularization methods is not driven by uncritical acceptance.

Even though we were able to give some valuable insights into the reasons why practitioners use regularizations, our study still has some limitations.
We have introduced regularizations through two examples referring to \textsc{lasso} models and mixed effects regression, which do not represent the variety of methods of regularizations in full.
This might have influenced the answers of the participants of the study, especially in context of the manipulation.
While we identified experience (\textsc{ex}) as a predictor of behavioural intention, our notion of experience was rather broad and did not include formal criteria such as statistical education of the survey participants.
A more formal approach towards experience might enable stratification for various levels of expertise, and therefore, could lead to more detailed insights into the connection between experience, behavioural intention, and other attitudes towards regularization.
Another key limitation is that there is no comparable research on the use and adaption of statistical methods.
Even though the application of the \textsc{utaut} model to our framework is carefully considered, it remains experimental here.
Furthermore, as the constructs trust and vigilance have been introduced only recently in this context, there is limited empirical evidence regarding their utility for explaining behaviour.
Therefore, we could not put our findings in a broader context. 

\section*{CRediT Author Statement}
K.E.T.: Methodology, Software, Formal analysis, Data Curation, Writing - Original Draft, Visualization, Project administration;
M.B.: Methodology, Software, Formal analysis, Investigation, Resources, Writing - Original Draft, Visualization, Project administration;
N.K.: Conceptualization, Methodology, Writing - Review \& Editing, Supervision, Funding acquisition;
A.G.: Methodology, Writing - Review \& Editing, Supervision;
M.P.: Conceptualization, Methodology, Writing - Review \& Editing, Supervision, Funding acquisition;
M.W.: Conceptualization, Methodology, Investigation, Resources, Writing - Original Draft.

\section*{Acknowledgement}
We thank Stefan Inerle for his helpful advice in handling psychometric data.
A.G.'s research was partially funded in the course of TRR 391 ``Spatio-temporal Statistics for the Transition of Energy and Transport'' (520388526) by the Deutsche Forschungsgemeinschaft (DFG, German Research Foundation).

\section*{Data Availability Statement}
The raw dataset and the data analysis scripts are openly available on TuDoData at \url{https://doi.org/10.17877/RCTRUST-2026-JP3FQ6}.

\printbibliography	
	
\newpage
\appendix

\end{document}